\documentclass[aps,prl,twocolumn,showpacs,amsmath,amssymb,floatfix,superscriptaddress]{revtex4-1}
\usepackage[dvipsnames, svgnames, x11names]{xcolor} 
\usepackage{graphicx}
\usepackage{dcolumn}
\usepackage{bm}
\usepackage{color}
\usepackage{graphicx}
\usepackage{epstopdf}
\usepackage{epsfig}
\usepackage[colorlinks=true, citecolor=black, anchorcolor=green]{hyperref}

\include{graphic}

\makeatletter
\begin{document}
\preprint{APS/123-QED}

\title{Unveiling a Pump-Induced Magnon Mode via its Strong Interaction with Walker Modes}

\author{J. W.~Rao}
\affiliation{School of Physical Science and Technology, ShanghaiTech University, Shanghai 201210, China}

\author{Bimu~Yao}\email{yaobimu@mail.sitp.ac.cn;}
\affiliation{School of Physical Science and Technology, ShanghaiTech University, Shanghai 201210, China}
\affiliation{State Key Laboratory of Infrared Physics, Shanghai Institute of Technical Physics, Chinese Academy of Sciences, Shanghai 200083, China}

\author{C. Y.~Wang}
\affiliation{School of Physical Science and Technology, ShanghaiTech University, Shanghai 201210, China}

\author{C.~Zhang}
\affiliation{School of Physical Science and Technology, ShanghaiTech University, Shanghai 201210, China}

\author{Tao~Yu}\email{taoyuphy@hust.edu.cn;}
\affiliation{School of Physics, Huazhong University of Science and Technology, Wuhan, 430074, China}

\author{Wei~Lu}\email{luwei@mail.sitp.ac.cn;}
\affiliation{School of Physical Science and Technology, ShanghaiTech University, Shanghai 201210, China}
\affiliation{State Key Laboratory of Infrared Physics, Shanghai Institute of Technical Physics, Chinese Academy of Sciences, Shanghai 200083, China}

\begin{abstract}

We observe a power-dependent anticrossing of Walker spin-wave modes under microwave pumping when a ferrimagnet is placed in a microwave waveguide that does not support any discrete photon mode. We interpret this unexpected anticrossing as the generation of a pump-induced magnon mode that couples strongly to the Walker modes of the ferrimagnet. This anticrossing inherits an excellent tunability from the pump, which allows us to control the anticrossing via the pump power, frequency, and waveform. Further, we realize a remarkable functionality of this anticrossing, namely, a microwave frequency comb, in terms of the nonlinear interaction that mixes the pump and probe frequencies. Such a frequency comb originates from the magnetic dynamics and thereby does not suffer from the charge noise. The unveiled hybrid magnonics driven away from its equilibrium enriches the utilization of anticrossing for coherent information processing.

\end{abstract}

\maketitle

\textit{Introduction}.---Anticrossing between two coupled bands is a universal phenomenon widely existing in a variety of systems, such as molecules \cite{lidzey1998strong,meerts1981avoided,benthem1978study}, hybrid quantum systems \cite{bylinkin2021real,lachance2020entanglement,fan2018superconducting,koh2012pulse}, electric circuits \cite{pozar2011microwave}, and mechanical devices \cite{verhagen2012quantum,pippard2007physics}. It leads to the formation of polaritonic quasiparticles, provides the foundation for coherent energy transfer between distinct physical entities   \cite{huang1951lattice,deng2010exciton,byrnes2014exciton,shen2015laser,kamra2015coherent,hayashi2018spin}, and has a strong effect on quantum transport. Anticrossing can be produced from either the drive field or the vacuum field. Specifically, driving a $\Lambda$-configuration system can lead to the anticrossing with the Autler-Townes (AT) effect \cite{autler1955stark,sillanpaa2009autler,saglamyurek2018coherent}. Its splitting gap is proportional to the drive field strength and is invaluable for accurate electromagnetic detection. On the other hand, using a cavity to reduce the photon mode volume can greatly enhance the light-matter interaction and thus produce the anticrossing, for instance the anticrossing in cavity magnonics \cite{rameshti2021cavity,soykal2010strong,huebl2013high,tabuchi2015coherent,bai2017cavity,goryachev2014high,lachance2019hybrid,bhoi2020roadmap,cao2015exchange,boventer2018complex,yu2019prediction,grigoryan2018synchronized,rao2021interferometric,li2022coherent} caused by the strong photon-magnon coupling. The recent work \cite{xu2020floquet} combines cavity magnonics with Floquet physics and enriches the control of magnon-based anticrossing by taking advantage of nonlinearity and non-equilibrium. However, it relies on coupling magnons to a cavity to constitute the $\Lambda$-configuration for AT splitting. It is a basic fact that on replacing the cavity by a waveguide, the discrete photon mode becomes a continuum and the magnon-based anticrossing vanishes.

\begin{figure} [h!]
\begin{center}
\epsfig{file=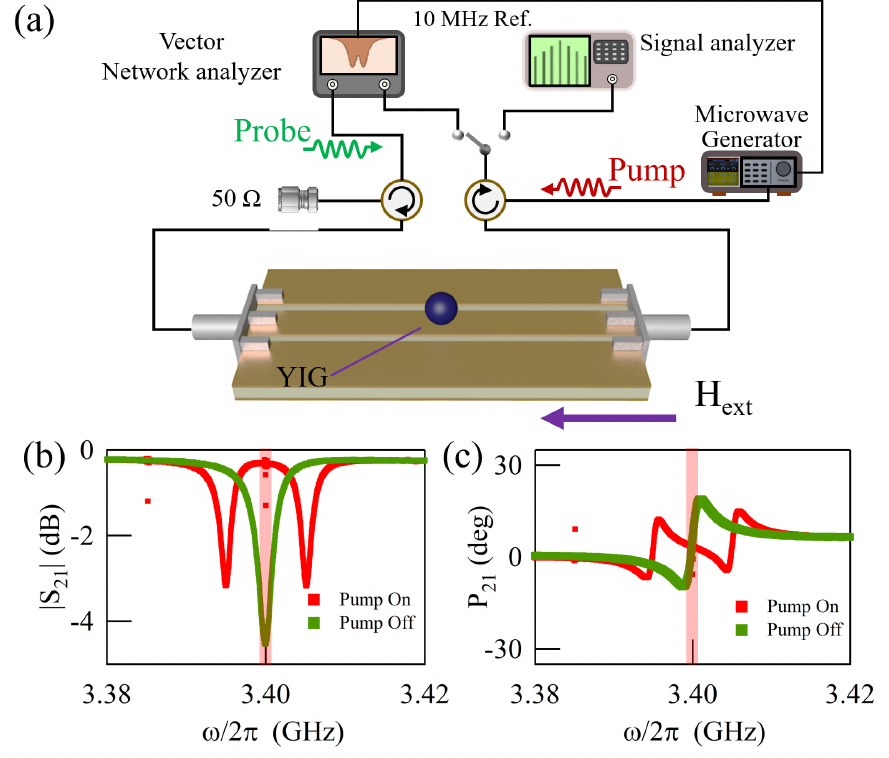,width=8.7cm} \caption{(a) Experimental configuration. Continuous pump and probe microwaves excite and detect the magnetization dynamics of a YIG sphere placed on top of a CPW. (b) and (c) Measured transmission magnitude ($|S_{21}|$) and phase ($P_{21}$) of the Walker $(2,2,0)$ mode. The red and green squares correspond respectively to the situations with and without the pump (5~dBm). Mode splitting appears after turning on the pump. The scattered data points at 3.384 and 3.4~GHz in both (b) and (c) are spurious signals arising from crosstalk between the pump and probe fields.}\label{fig1}
\end{center}
\end{figure}

In this work, however, we observe an anticrossing of Walker modes \cite{walker1957magnetostatic,dillon1957ferrimagnetic,gloppe2019resonant} during strong driving of a ferrimagnetic yttrium iron garnet (YIG) sphere on top of a coplanar waveguide (CPW) by microwaves, as illustrated in Fig.~\ref{fig1}(a). This waveguide is a broadband microwave device that has no discrete photon modes, and therefore anticrossing is not expected \cite{yu2020chiral,yu2020magnon}. The anticrossing, on the other hand, is not an intrinsic property of the system but is tunable by the pump power $P_d$ with the gap scaling as $P_d^{1/4}$, which makes strong distinction from that of AT splitting $\propto P_d^{1/2}$. As a phenomenological theory, we interpret the other discrete mode, which couples strongly to the Walker mode to induce the anticrossing, as a pump-induced magnon mode (PIM) \cite{rameshti2021cavity,bryant1988spin,SM}. The PIM can be continuously tuned by the pump up to a saturated frequency, while maintaining a narrow linewidth ($\sim$ 500 kHz). When the anticrossing appears, the radiation of our system reads out multiple peaks with equal frequency intervals, thus demonstrating a frequency comb \cite{wang2021magnonic,hula2021spin,schliesser2012mid,kippenberg2011microresonator,coddington2008coherent} functionality, produced by mixing the pump and probe signals via the nonlinear magnon interaction, including cross-Kerr effect \cite{wu2021observation,hoi2013giant,sheng2008efficient}. 

\textit{Anticrossing of Walker modes in a waveguide}.---The CPW, fabricated on a $22\times50$~mm$^2$ RO4350B board, has a standard impedance of 50 $\Omega$. The width of its central strip (1.1 mm) is slightly larger than the diameter of the YIG sphere (1 mm). The gap between the central strip and the two ground planes is 0.24 mm. The YIG sphere placed on top of the central strip of the CPW has a number of spin $N_w\approx10^{19}$ and saturated magnetization \cite{elyasi2020resources} $M=1.46\times 10^5$~A/m [Fig.~\ref{fig1}(a)]. An external magnetic field $H_{\rm ext}$ is applied to tune the Walker mode frequency $\omega_w$. Two continuous microwaves, namely, the pump and probe, respectively drive and detect the ferromagnetic resonance. The pump, restricted to a single frequency $\omega_d$, has a large power $P_d$. The probe, produced and received by a vector network analyzer (VNA) for transmission ($S_{21}$) measurement, has a fixed power of $-25$ dBm throughout the experiment, but sweeps a wide-band frequency $\omega_p$. A signal analyzer (SA) monitors the radiation spectra of the system, enabling the study of frequency conversion.

Without the pump, we observe several Walker modes of the YIG sphere from the transmission spectrum \cite{SM}. As analytical solutions of the Maxwell and Landau-Lifshitz equations with spherical boundary conditions, Walker modes can be simply viewed as standing spin waves in the YIG sphere \cite{walker1957magnetostatic,dillon1957ferrimagnetic,gloppe2019resonant}. Here, we focus on the  $(2,2,0)$  Walker mode. Its transmission amplitudes and phases are plotted as green squares in Figs.~\ref{fig1}(b) and (c), respectively. They can be reproduced well by a standard transmission formula $S_{21}=1+\kappa/[i(\omega-\omega_w)-(\kappa+\alpha)]$ with $\omega_w/2\pi=3.4$~GHz for the Walker mode frequency at 116.5~mT, an external damping rate $\kappa/2\pi=0.55$~MHz, and an intrinsic damping rate $\alpha/2\pi=0.85$~MHz.

After we turn on the pump with frequency $\omega_d/2\pi=3.4$~GHz and power $P_d=5$~dBm, the former single resonance dramatically splits into two dips with equal intensities, shown by the red squares in Fig.~\ref{fig1}(b). The pump-induced splitting (PIS) gap is 10 MHz, much larger than the Walker mode linewidth $(\kappa+\alpha)/2\pi=1.4$~MHz. At each resonant dip, there exists a phase jump of the microwaves, as shown in Fig.~\ref{fig1}(c).

\begin{figure} [htbp]
\begin{center}
\epsfig{file=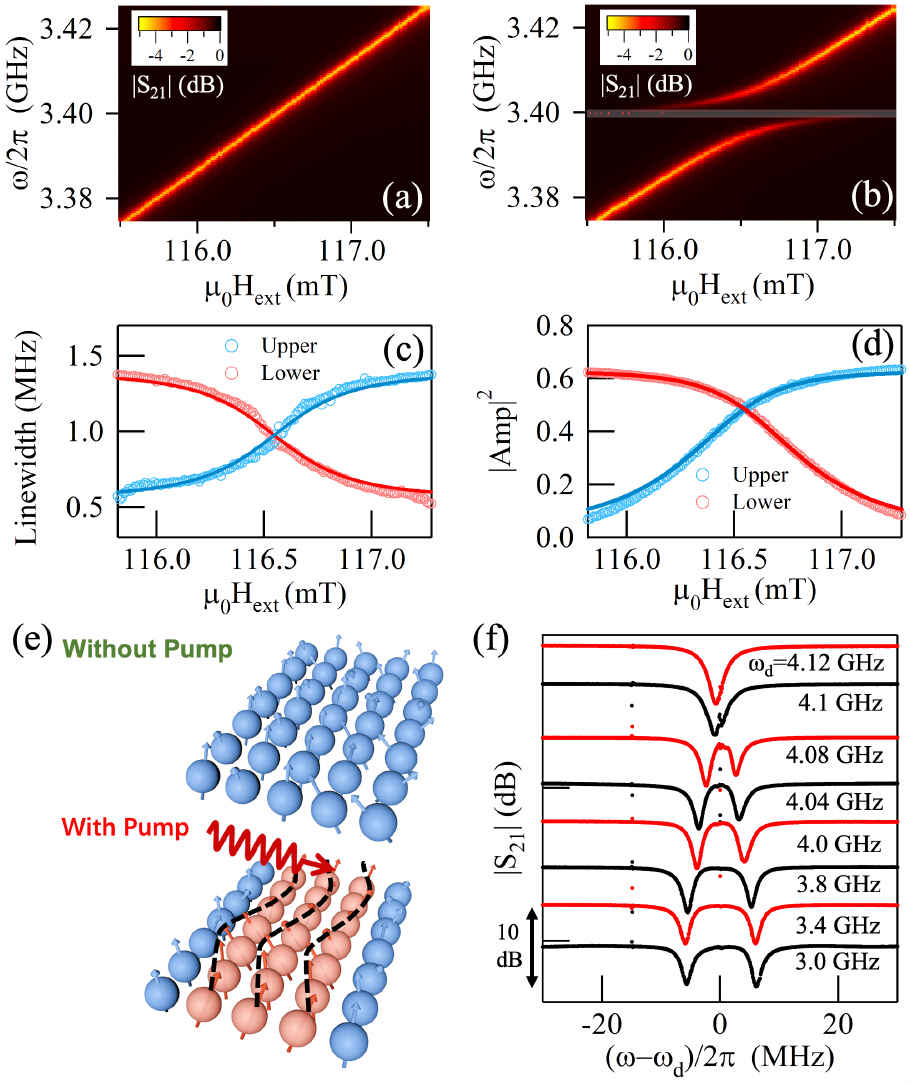,width=8.6cm} \caption{ (a) and (b) Transmission spectra of the Walker mode measured at different $H_{\rm ext}$ without and with the pump. (c) and (d) Linewidths and squared transmission amplitudes of the upper (blue circles) and lower (red circles) branches in (b) as functions  of $H_{\rm ext}$.  The solid lines are the results calculated using Eq.~(\ref{Hamiltonian}). (e) The cooperative precession of unsaturated spins under pump as possible origin of PIM. (f) Evolution of PIS when increasing $\omega_d$.}\label{fig2}
\end{center}
\end{figure}

The PIS also exhibits a nontrivial dependence on the external magnetic field $H_{\rm ext}$. Without the pump, a linear dependence of the Walker mode ($\omega_w$) on $H_{\rm ext}$ is shown in Fig.~\ref{fig2}(a), following the Kittel formula \cite{walker1957magnetostatic}. We then turn on the pump and perform the measurement with the same experimental settings. The former linear dispersion is split into two branches with an anticrossing  [Fig.~\ref{fig2}(b)]. This reveals a strong coupling between the Walker mode and a hidden discrete mode, which oscillates at the pump frequency and is insensitive to any $H_{\rm ext}$. The two branches follow a dispersion $\big[\omega_d+\omega_w\pm \sqrt{(\omega_d-\omega_w)^2+4g^2}\big]/2$ with a coupling strength $g/2\pi=5$~MHz. Their linewidths and amplitudes, extracted from Fig.~\ref{fig2}(b), are plotted in Figs.~\ref{fig2}(c) and (d), respectively. With increasing $H_{\rm ext}$, the linewidths of the two branches exchange values between 1.4 and 0.55~MHz, while their squared transmission amplitudes exchange values between 0.6 and nearly zero. These features in Figs.~\ref{fig2}(b)-(d) are signatures of a strong coupling effect between two discrete modes. Although the pump-induced mode is invisible in the transmission, the amplitude and linewidth evolution of the anticrossing can help us to reveal its dynamic properties.

The PIS is not restricted to the $(2,2,0)$ mode, but is a general feature of other Walker modes \cite{SM}. It cannot be explained by presently known nonlinear mechanisms, including the self-Kerr effect \cite{wang2018bistability,zhang2019theory}, the Suhl instability \cite{anderson1955instability,elyasi2020resources}, and the spin nutation \cite{PhysRevX.9.041036}.  Moreover, when we replace the CPW with a different microwave device, such as a planar cavity, a microstripline, or a loop antenna, the PIS is still present. Therefore, we exclude the possibility that PIS originates from specific electromagnetic surroundings.

Based on existing knowledge, the known magnon modes in a YIG sphere, including Walker and Damon-Eshbach (DE) modes, have no resonance matching the pump frequency that is independent of the magnetic field. We conjecture that, at the low magnetic field, there exist many unsaturated spins in the YIG with random precession. After applying a pump, such unsaturated spins may obtain cooperativity and form a spin wave [Fig.~\ref{fig2}(e)]. We formulate a phenomenological model to account for this spin wave, namely the PIM, but leave the microscopic details to the fitting parameters \cite{SM}. To support our assumption, we measure the PIS at different  $\omega_d$ [Fig.~\ref{fig2}(f)]. With increasing $\omega_d$ to large $H_{\rm ext}$, the YIG gradually saturates. The measured PIS closure in Fig.~\ref{fig2}(f) indicates the disappearance of PIM. Considering possible disturbance from thermal fluctuations on unsaturated spins, we envision exploring the temperature dependence of PIS, especially at the millikelvin temperatures, can provide more in-depth clues to its microscopic origin.

The PIM's radiation, however, is obscured by the pump, making its direct observation difficult. However, via its interaction with the Walker modes, we can detect it indirectly. We mainly consider the Zeeman interaction between two modes, which is modelled as $F_{\rm in}=\eta\mu_0\gamma\int_{V}\textbf{M}_w({\bf r})\cdot \textbf{M}_p({\bf r})d{\bf r}$ \cite{zhang2019theory,SM}, where $\eta$ represents an effective demagnetization factor $O(1)$, $\textbf{M}_w({\bf r})$ and $\textbf{M}_p({\bf r})$ are the magnetization of two modes, $\gamma$ is the electronic gyromagnetic ratio and $V$ is the YIG volume \cite{wu2021observation}. Note that we have ignored the spin textures of two modes in this effective model, and focus only on the key features of our nontrivial observations.  

We denote by $\hat{a}^\dag(\hat{a})$ and $\hat{b}^\dag(\hat{b})$, the creation (annihilation) operators of the Walker mode and the PIM, respectively, which couple in the Hamiltonian
\begin{align}
&\mathcal{H}/\hbar=\tilde{\omega}_w\hat{a}^\dag
\hat{a} +\tilde{\omega}_d\hat{b}^\dag\hat{b}+ g(\hat{a}^\dag\hat{b}+\hat{a}\hat{b}^\dag)+ K\hat{a}^\dag\hat{a}\hat{b}^\dag\hat{b}+\Omega_p\nonumber\\
&\quad(\hat{a}^\dag e^{-i\omega_pt}+{\rm H.c.})+((\Omega_d\hat{a}^\dag+\Omega'_d\hat{b}^\dag) e^{-i\omega_dt}+{\rm H.c.}), 
\label{Hamiltonian}
\end{align}
where $\tilde{\omega}_w=\omega_w-i(\kappa+\alpha)$ and $\tilde{\omega}_d=\omega_d-i\xi$ are the complex mode frequencies. The PIM's damping rate is $\xi/2\pi=0.55$ MHz.  $\Omega_p$, $\Omega_d$, and  $\Omega'_d$ correspond to the effective intensities of the probe and pump felt by each mode. Two modes' interaction includes: (i) a coherent coupling with a strength $g=\eta\mu_0\hbar^2\gamma^3\sqrt{N_wN_p}/\sqrt{2}V$, where $N_p$ represents the PIM's spin number, which should be the magnon number excited by the pump, i.e., $N_p=\langle\hat{b}^\dag\hat{b}\rangle$. (ii) a cross-Kerr effect with a coefficient $K=g/\sqrt{N_wN_p}$. Since $N_w,N_p\gg1$, we have $g\gg K$ \cite{SM}. This weak $K$ does not induce obvious frequency shifts for the two hybrid modes, but can lead to frequency conversion, i.e., a portion of system energy is converted to sidebands and generates new magnons. This process is beyond the measurement capability of the VNA, and will be addressed in the frequency comb measurement later.

Driven by the pump, our system mainly oscillates at $\omega_d$. In addition, small fluctuations are induced by the weak probe \cite{SM}. Expressing both the steady state and fluctuations in terms of the Walker mode and PIM, we derive the coupling strength \cite{SM}
\begin{equation}
g=\sqrt{\sqrt{g_0^2\kappa A_d^2-(\xi-\varrho A_d)^2\Delta^2}-(\xi-\varrho A_d)(\kappa+\alpha)},
\label{CS}
\end{equation}
where $A_d$ is the pump field amplitude, $\Delta=\omega_w-\omega_d$ is the detuning between two modes, and $g_0$ and $\varrho$ are two constants \cite{SM}. We further linearize the response of the probe and obtain the transmission spectrum
\begin{equation}
S_{21}\approx1+\frac{\kappa}{i(\omega_p-\omega_w)-(\kappa+\alpha)+\frac{g^2}{i(\omega_p-\omega_d)-\xi}},
\label{Transmission}
\end{equation}
from which the calculated linewidths and amplitudes of the two hybridized modes agree with the experimental findings in Figs.~\ref{fig2}(c) and (d).

\begin{figure} [ht]
\begin{center}
\epsfig{file=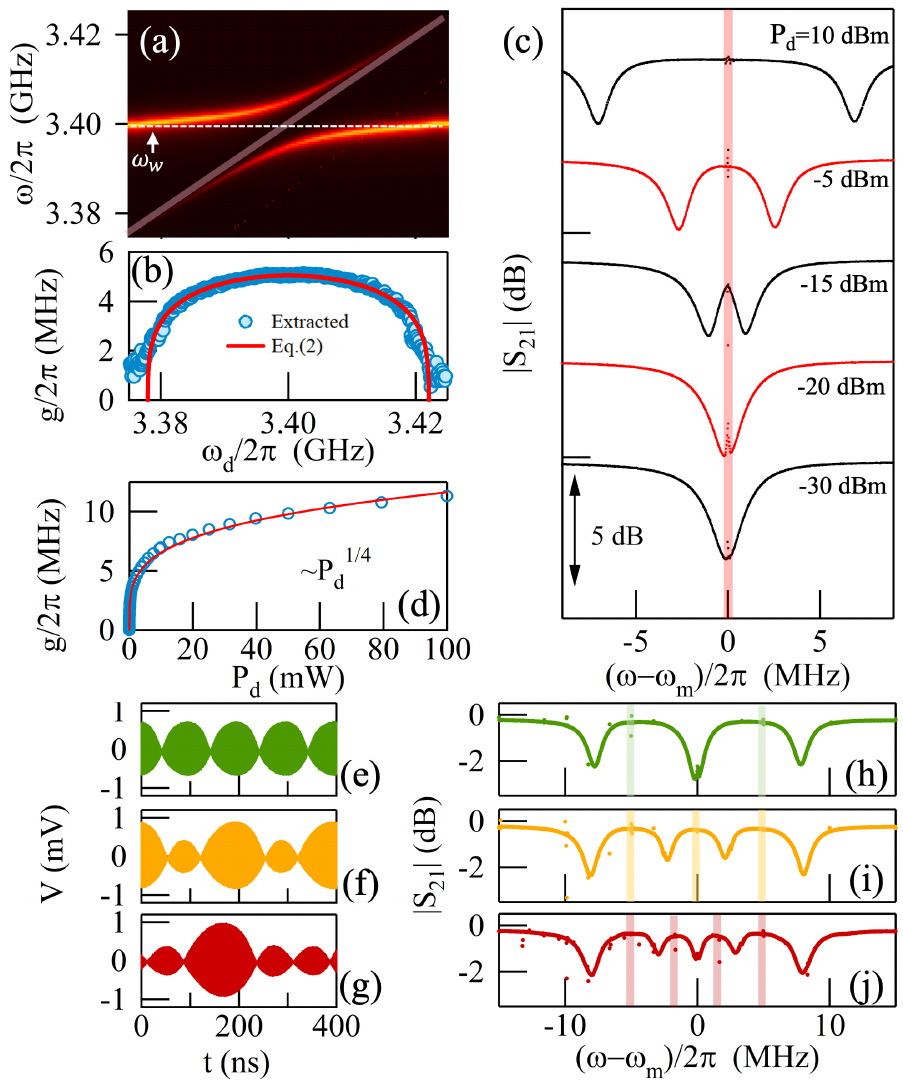,width=8.5 cm} \caption{(a) and (b) Measured PIS transmission and extracted coupling strength $g$ as function of $\omega_d$. The red line is calculated by using Eq.~(\ref{CS}). (c) Power dependence of the PIS. The pump is indicated by the pink strips in (a) and (c). (d) Coupling strength $g$ as a function of $P_d$, which is well fitted by $g\propto P_d^{1/4}$ by the red line. (e)-(g) Synthetic pumps with time-dependent amplitudes, consisting of $N=2$, 3, or 4 main tones with equal intensities, constant phases and equal frequency intervals represented by the colored strips in (h)-(j), where three, four, and five hybridized magnon modes appear. }\label{fig3}
\end{center}
\end{figure}

\textit{Pump dependence}.---We now study the dependence of the anticrossing on the pump frequency $\omega_d$. Figure~\ref{fig3}(a) shows the transmission spectra measured at different $\omega_d$ but a constant power $P_d=5$~dBm. With constant $H_{\rm ext}$, the Walker mode $\omega_w/2\pi$ is fixed at 3.4 GHz, but the PIM can be continuously tuned by varying $\omega_d$. Hence, we tune the PIM to cross the Walker mode and thereby produce an anticrossing [Fig.~\ref{fig3}(a)]. The coupling strength at different $\omega_d$ are extracted \cite{SM} from Fig.~\ref{fig3}(a) and plotted as blue circles in Fig.~\ref{fig3}(b). It varies slowly near resonance, but drops quickly at large detuning. This variation is well reproduced by Eq.~(\ref{CS}), shown as the red line in Fig.~\ref{fig3}(b).

We then address the power dependence of the anticrossing, which shows that this effect is not an intrinsic property like the cavity magnon polariton \cite{rameshti2021cavity,soykal2010strong,huebl2013high,tabuchi2015coherent,bai2017cavity,goryachev2014high,lachance2019hybrid,bhoi2020roadmap,cao2015exchange,boventer2018complex,yu2019prediction,grigoryan2018synchronized,rao2021interferometric,li2022coherent}, but an effect induced by the pump. To this end, we set $\omega_d/2\pi=3.4$~GHz and vary $P_d$ from $-30$ to 20~dBm. With a low power ($-30$~dBm) we only observe a single dip, which splits into a doublet at the power around $-20$~dBm, as shown in Fig.~\ref{fig3}(c). As $P_d$ increases further, the splitting gap increases significantly. Figure~\ref{fig3}(d) shows the coupling strength $g$, extracted from each spectrum, as a function of $P_d$. Through fitting we find the scaling relation $g/2\pi=C\cdot  P_d^{1/4}$, with just one fitting parameter $C=3.7$~$\rm MHz/\sqrt[4]{\rm mW}$. This relation is again reproduced by Eq. (\ref{CS}), since $A_d\propto\sqrt{P_d}$.

The PIM-Walker mode coupling can also be controlled by modulating the pump waveform. We replace the single-frequency pump with synthetic pumps consisting of $N=2$, 3, or 4 main tones with equal intensities, constant phases, and equal frequency intervals. Their amplitudes are time-dependent, as shown in Figs.~\ref{fig3}(e)-(g). The corresponding main tones are indicated by the vertical strips in Figs.~\ref{fig3}(h)-(j). Each tone generates a PIM, which then couples with the Walker mode, resulting in $N+1$ hybridized modes. This approach of generating multiple hybridized magnon modes relies simply on the waveform manipulation, which thus may find applications in magnon-based coherent information storage \cite{zhang2015magnon} or Floquet electromagnonics \cite{xu2020floquet}.

\begin{figure} [ht]
\begin{center}
\epsfig{file=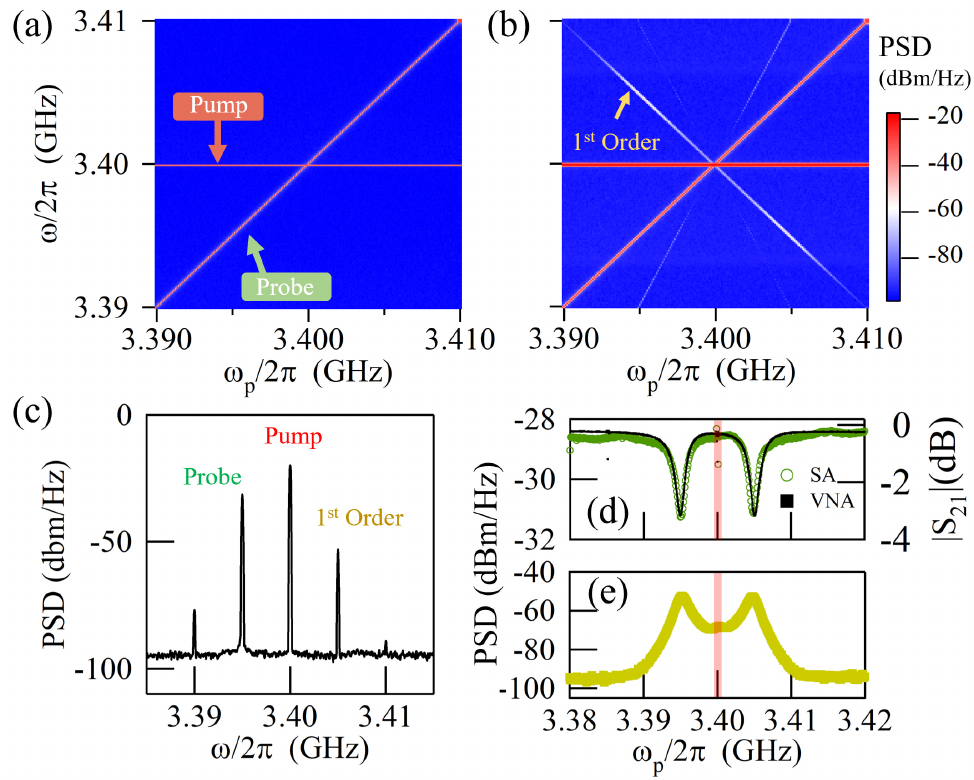,width=8.6cm} 
\caption{(a) and (b) Measured radiation spectra at $H_{\rm ext}=0$ and $\omega_w/2\pi=3.4$ GHz under $\omega_d/2\pi=3.4$ GHz. The probe frequency $\omega_p$ is varied from 3.39 to 3.41~GHz. (c) Frequency comb measured at $\omega_p/2\pi=3.395$~GHz. (d) Amplitude of the probe extracted from (b) (green circles) as a function of $\omega_p$. Two hybridized magnon modes are consistent with the VNA measurement (black squares). (e) Amplitude of the first-order sideband signal as a function of $\omega_p$, which reaches the maximum at two hybridized modes.}\label{fig4}
\end{center}
\end{figure}

\textit{Frequency comb}.---Besides the linear coupling, the nonlinear PIM-Walker mode interaction enables frequency conversion. We set the pump as $\omega_d/2\pi=3.4$ GHz and $P_d=5$ dBm. The probe frequency $\omega_p$ is swept from 3.39 to 3.41 GHz successively. At each $\omega_p$, the radiation spectrum of our system, i.e., the power spectral density (PSD), is recorded by the SA. Figure~\ref{fig4}(a) shows the measured results at zero $H_{\rm ext}$, i.e., $\omega_w=0$. Two signals simply cross each other, and no sidebands appear. We then tune the Walker mode to match the pump frequency, i.e., $\omega_w/2\pi=\omega_d/2\pi=3.4$~GHz, and keep all other settings unchanged. From the measured PSD [Fig.~\ref{fig4}(b)], we observe a frequency comb. Beside the pump and probe, the first-, second- and third-order sideband signals occur at $(2\omega_d-\omega_p)/2\pi$, $(2\omega_p-\omega_d)/2\pi$, and $(3\omega_d-2\omega_p)/2\pi$, respectively. Figure~\ref{fig4}(c) shows a spectrum measured at $\omega_p/2\pi=3.395$ GHz. The sideband signals are several orders of magnitude weaker than the detected pump and probe.

The amplitudes of all the comb teeth change with the probe frequency $\omega_p$. Here we extract the amplitudes of the probe and the first-order sideband signal from Fig.~\ref{fig4} (b) and plot them in Figs.~\ref{fig4}(d) and (e). From the probe [green dots in Fig.~\ref{fig4}(d)], we see two hybridized modes at 3.395 GHz and 3.405 GHz, consistent with the transmission measurement (black line). These two modes are two absorption channels for the probe, at which the energy absorption and conversion is most efficient. Consequently, the first-order sideband signal reaches its maximum [Fig.~\ref{fig4}(e)]. Based on the Hamiltonian (\ref{Hamiltonian}), we numerically simulate the frequency comb arising from the nonlinear magnon interaction, including the cross-Kerr effect, which reproduces our experimental observations well \cite{SM}. 

\textit{Conclusion and discussion}.---We have observed an anticrossing of Walker modes, when a YIG sphere is driven by microwaves in a waveguide. This phenomenon resembles a transcritical bifurcation in nonlinear physics \cite{crawford1991introduction}, but has a different physical origin. If we assume that a driven magnon mode, i.e., the PIM, is induced by the pump, we can naturally reproduce the key features of our observations. The PIM is not directly observed with the pump-probe technique, but is traced via its interactions with the Walker modes. 

Our study demonstrates that a magnet driven away from equilibrium can exhibit interesting and potentially useful properties \cite{de2021colloquium}. The pump-induced anticrossing in our work has a cooperativity higher than 120, unambiguously demonstrating the strong coupling effect. This anticrossing exhibits flexible controllability by the pump frequency, power and waveform. Such an excellent flexibility is rarely achievable with cavity magnon polaritons. Although performing below a saturated frequency, the PIM operates robustly in the S-band that is vital to wireless networking\cite{sharma2021electrically} and quantum information processing\cite{amsuss2011cavity}.

The PIM also enables the mixing of the pump and probe and hence generates a frequency comb. This frequency comb is constructed purely on the basis of magnetic nonlinearity and hence does not suffer from the charge noise from electric nonlinearities \cite{pozar2011microwave}. Utilizing our frequency comb, we may be able to realize coherent information conversion with ultralow noise or study spin-wave solitons in magnonics. After lifting a corner of the veil covering the PIM, we will further investigate more of the nature and functionality of this nonequilibrium mode.

\vskip0.25cm 
\begin{acknowledgments}
This work has been funded by National Natural Science Foundation of China under Grants Nos.11974369, 12122413, 12204306 and \textcolor{black}{0214012051}, STCSM Nos.21JC1406200 \textcolor{black}{and 22JC1403300}, the Youth Innovation Promotion Association  No.~2020247 \textcolor{black}{and Strategic priority reasearch No. XDB43010200} of CAS, the SITP \textcolor{black}{Independent} Foundation, the Shanghai Pujiang Program (No. 22PJ1410700) and the startup grant of Huazhong University of Science and Technology (Grants Nos.3004012185 and 3004012198).
\end{acknowledgments}

\end{document}